\documentclass[12pt,a4paper,oneside]{article}
\usepackage{graphicx,sectsty,amssymb,amsmath,subfig}
\usepackage[linkcolor={blue},citecolor={red},colorlinks=true]{hyperref}
\usepackage[margin=2cm]{geometry}
\begin{document}

 \title{\bf Spin Determination via Third Generation Cascade Decays}
 \author{Oram Gedalia$^a$, Seung J.~Lee$^a$ and Gilad Perez$^{a,b}$ \vspace{6pt}\\
 \fontsize{10}{16}\selectfont\textit{$^a$ Department of Particle Physics, Weizmann Institute of Science, Rehovot 76100, Israel} \vspace{2pt}\\
 \fontsize{10}{14}\selectfont\textit{$^b$ C.~N.~Yang Institute for Theoretical Physics, Stony Brook University,}\\
 \fontsize{10}{14}\selectfont\textit{Stony Brook, NY 11794-3840, USA}}
 \date{}
 \maketitle

\begin{abstract}
Once new particles are discovered at the LHC and their masses are
measured, it will be of crucial importance to determine their
spin, in order to identify the underlying new physics model. We
investigate the method first suggested by Barr and later extended
by others to distinguish between Supersymmetry and alternative
models, e.g.~Universal Extra Dimensions, in a certain cascade
decay. This method uses invariant mass distributions of the
outgoing Standard Model particles to measure the spin of
intermediate particles, by exploiting the quark/anti-quark
asymmetry of the LHC as a $pp$ collider, which is limited for
first generation quarks. In this work, we suggest instead to
measure the charge of the outgoing quark, in case it is a third
generation quark. The resulting asymmetry for a bottom quark is
similar to the previous method, while it is independent of
hadronic uncertainties. Furthermore, for a top quark, the
asymmetry allows better distinction between the models, as
demonstrated by a quantitative analysis of model discrimination.
We also show that the top's decay products can be used instead of
the top itself, when the reconstruction of the top momentum is
difficult to accomplish, and still provide information about the
spin.
\end{abstract}

 \vskip 1.3cm

\section{Introduction}

The Large Hadron Collider (LHC), which has already started and
will resume operating soon, is expected to unveil new physics at
the TeV scale. If we assume that the underlying new physics is
weakly coupled at the TeV scale, and also assume the naturalness
paradigm for the light Higgs, then we expect that there will be
new particle sectors in order to cancel quadratic divergencies
appearing in the Standard Model (SM) loop contributions to the
Higgs mass. In particular, since the largest contribution arises
from loop involving top quark, we expect that there will be a new
particle (often called top-partner) cancelling the divergence from
the top. Since the top and the bottom are in the same SU(2)
multiplet, it is interesting to look for new physics partners for
third generation quarks.

The most well-known example is low energy Supersymmetry (SUSY),
where the quadratic divergence is cancelled by its scalar partner,
with its spin differing by $1/2$. On the other hand, there is a
different class of new physics models, such as Universal Extra
Dimensions (UED)~\cite{ued} and Little Higgs
models~\cite{ArkaniHamed:2001ca,ArkaniHamed:2001nc}, where the
divergences are cancelled by partner particles with the same spin.
Furthermore, these models often contain a discrete symmetry, such
as R-parity for SUSY, KK-parity for UED (and a certain variation
of warped extra dimension model \cite{Agashe:2007jb}) and T-Parity
\cite{t_parity} for Little Higgs models, so that the lightest
parity-odd particle becomes a natural dark matter candidate, which
will presumably show up as missing energy in collider experiments
at the end of cascade decays. According to all these scenarios and
many more, new particles should be observed at the LHC, typically
after decaying to SM particles.

However, once we discover new particles, it will not be sufficient
to know the mass spectrum, production cross sections, or decay
branching ratios, if we wish to understand what the underlying new
physics is. Determining the spin of these newly-discovered
particles is crucial in order to distinguish among new physics
models, in particular, between SUSY (where partners of SM
particles have spin differing by $1/2$) and same-spin theories
(which we will refer to as ``UED'' from now on as an illustrative
name, since these theories are similar for our purpose).

Various methods for spin measurements have been suggested and
discussed before
\cite{barr,smillie_old,smillie,Battaglia:2005zf,Datta:2005zs,
Choi:2006mr,Choi:2006mt,Meade:2006dw,Wang:2006hk,Kilic:2007zk,Alves:2007xt,
Rajaraman:2007ae,Csaki:2007xm,Kane:2008kw,Wang:2008sw,Buckley:2008eb,
Burns:2008cp,Cho:2008tj,probing}. In this work, we focus on the
following decay chain within the Minimal Supersymmetric Standard
Model:
\begin{equation} \label{decay_chain}
\tilde{q} \to q \tilde{\chi}^0_2 \to q l^{\pm} \tilde{l}^{\mp} \to q
l^{\pm} l^{\mp} \tilde{\chi}^0_1 \, ,
\end{equation}
where the spins of the intermediate particles need to be
determined in order to establish the correct model behind the
chain. In this process the $\tilde{\chi}^0_1$, presumably the LSP,
escapes detection, so the spin measurement must be based on the
angular distribution of the quark jet and the two leptons.
Moreover, it seems very hard to distinguish between the first
emitted lepton (the \emph{near} lepton) and the second one (the
\emph{far} lepton) and to identify the charge of the quark (that
is, if it is $q$ or $\bar{q}$). Hence, the measured angular
distribution cannot be correctly assigned to the outgoing
particles.

The method first suggested by Barr \cite{barr}, and later
investigated by others \cite{smillie_old,smillie}, uses the
production asymmetry of the LHC as a $pp$ collider. Since the
colliding particles are both protons, there is more chance to get
a quark than an anti-quark. This asymmetry was estimated to be
around 70\%, based on the difference between the valence and the
sea quarks in the Parton Distribution Functions (PDFs).

Our suggestion to gain further information in this measurement,
independent of the PDFs, is to identify the charge of the
quark\footnote{This was also recently suggested in \cite{probing}.
The approach of that paper is more model-independent than ours, as
they characterize all the theoretical possibilities in the
process. However, they use distributions which are not directly
observable, while leaving the problem of distinguishing the near
and far leptons to future work.}. This is probably impractical for
a light quark jet, but it is possible for third generation quarks.
A similar idea was utilized for the spin determination of the
gluino \cite{eboli}. The charge ID of the bottom and the top was
recently investigated thoroughly in several different contexts
(see for example
\cite{single_top,top_charge,top_charge_asymmetry,cpviolation}).
The basic idea for measuring the charge is to use the leptonic
decay channel of the $W$, which is emitted from the heavy quark.

This paper is organized as follows: The basic inputs we need are
presented in section~\ref{mass_spectrum}; the theoretical angular
distributions required for the analysis are given in
section~\ref{Angular_Distributions};
section~\ref{exp_distributions} discusses the method for spin
determination using the distributions; in section~\ref{mc} we
support the calculations with Monte Carlo simulations; we propose
an approach to circumvent full event reconstruction in
section~\ref{daughters}; actual model discrimination based on our
method is discussed in section~\ref{discrimination}; we conclude
in section~\ref{conclusion}.

\section{Mass Spectrum and Cross Sections} \label{mass_spectrum}

The mass spectrum of the intermediate particles, which affects the
analysis discussed below, is of course model-dependent. Here we
consider one of the Snowmass points for the SUSY scenario (SPS1a)
\cite{snowmass,snowmass3} and one representing case for the UED
scenario\footnote{Both spectra were considered in \cite{smillie}
(the UED spectrum is based on \cite{kk_masses}), so they were
chosen here in order to allow for an apples-to-apples
comparison.}. The mass spectra for the SUSY and UED scenarios are
shown in tables \ref{susy_spec} and \ref{ued_spec}, respectively.
The particles in the decay process are denoted by A, B, C, and D
(see figure~\ref{fig:decay chain}). Note that we focus on third
generation squarks or same-spin quark-partners.

\begin{figure}[hbt]
\centering
\includegraphics[height=1.5In]{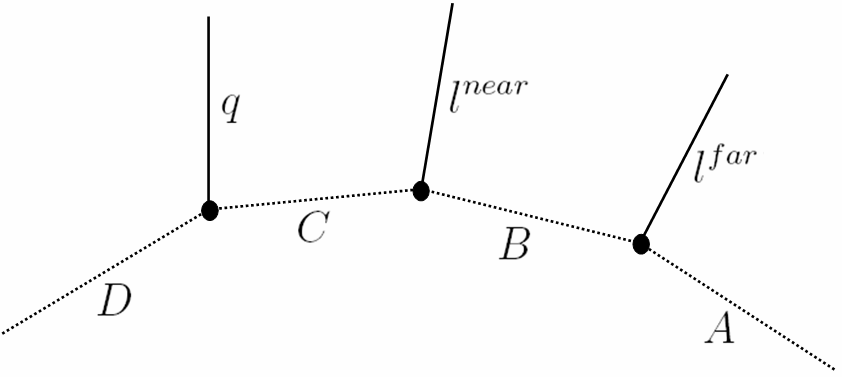}
\caption{The decay chain under consideration} \label{fig:decay
chain}
\end{figure}

\begin{table}[hbt]
\begin{center}
\begin{tabular}{|c|c|c|c|c|c|c|c|}
\hline \multicolumn{4}{|c|}{D} & C & \multicolumn{2}{|c|}{B} & A \\
\hline $\tilde{b}_1$ & $\tilde{b}_2$ & $\tilde{t}_1$ &
$\tilde{t}_2$ & $\tilde{\chi}^0_2$ & $\tilde{e}_L$ & $\tilde{e}_R$ & $\tilde{\chi}^0_1$ \\
\hline 507 & 540 & 391 & 583 & 179 & 202 & 144 & 97 \\
\hline
\end{tabular}
\end{center}
\caption{SUSY SPS1a mass spectrum (GeV)} \label{susy_spec}
\end{table}

\begin{table}[hbt]
\begin{center}
\begin{tabular}{|c|c|c|c|c|c|c|}
\hline \multicolumn{4}{|c|}{D} & C & B & A \\ \hline $t^*_1$ & $t^*_2$ & $b^*_1$ & $b^*_2$ & $Z^*$ & $l^*_L$ & $\gamma ^*$ \\
\hline 909 & 951 & 936 & 936 & 851 & 824 & 800 \\ \hline
\end{tabular}
\end{center}
\caption{UED mass spectrum (GeV)} \label{ued_spec}
\end{table}

The particles denoted by D in these two spectra, which are mass
eigenstates, are generically mixtures of the left- and
right-handed states. It turns out that for SPS1a, the stops and
the sbottoms are in fact large admixtures of the chiral
eigenstates, while for the UED spectrum considered here, the
heavier states are almost completely identical to the left-handed
(LH) states, that is, the mixing is very small.

For later convenience, as in \cite{smillie}, we define the mass
ratios
\begin{equation}
x=m^2_C/m^2_D \, , \qquad y=m^2_B/m^2_C \, , \qquad z=m^2_A/m^2_B \,
,
\end{equation}
so that $0 \leq x,y,z \leq 1$.

It should be noted that in the actual experiment, the mass of the
intermediate particles will be measured prior to the spin
determination, with uncertainties of about 5-10\%
\cite{mass_deter,:1999fr,Ball:2007zza}. Hence the analysis below
is based on the assumption that these masses are already known
(the corresponding uncertainties are neglected, for simplicity).

We have computed the cross sections for the processes under
consideration for the mass spectrum of SPS1a, using
MadGraph/MadEvent~\cite{Stelzer:1994ta,Maltoni:2002qb,madgraph}.
For example, the cross section for the production of
$\tilde{t}_1$, via direct processes and indirect processes (e.g.\
through gluino decay), is in the order of a few pb. Restricting
our attention to the decay chain \eqref{decay_chain}, reduces this
to a few tens of fb (we ignore tau leptons\footnote{Actually, some
preliminary work shows that taus might be tagged with significant
efficiency at the LHC~\cite{tau}. This has the potential of
greatly increasing our signal, since $\tilde{\chi}_2^0$ mostly
decays to staus (almost 90\% for SPS1a), relative to selectrons
and smuons.}). Another suppression by a factor of five stems from
the requirement for a semileptonic decay of the quark, to enable
the measurement of its charge. Overall, we expect that an
integrated luminosity of $\sim$100 fb$^{-1}$ should provide enough
statistics for the analysis. For the heavier stop state,
$\tilde{t}_2$, the production cross section is about an order of
magnitude lower. This is a result of the large mass difference
between the two states, and the fact that it is accidently almost
degenerate with the gluino, so that indirect production processes
are suppressed. Of course, the cross section is highly
model-dependent.

Comparing this to the case of first generation squarks, the total
production cross section of $\tilde{u}_L$ and $\tilde{d}_L$ is
about a factor of 2-3 higher than for $\tilde{t}_1$, the branching
ratio for this decay chain is a factor of 3 higher, and obviously
there is no dilution coming from charge measurement. Therefore,
the statistics should be about a factor of 40 higher than for
$\tilde{t}_1$.

A combinatorial confusion with the other side of the cascade decay
is also possible. If the other supersymmetric/same-spin partner
produced in the process decays through a similar chain, then it
would be extremely difficult to correctly assign all the outgoing
quarks and leptons. However, this is usually not the case, since
shorter decay chains (e.g.~$\tilde{t}_1 \rightarrow
\tilde{\chi}_1^+ b$) or chains involving taus and neutrinos are
much more likely to take place. Hence, in our work, we focus on
general cases where such an issue does not present a problem.

\section{Theoretical Angular Distributions} \label{Angular_Distributions}

For a fixed spin assignment, there are two possible angular
distributions within the chain, as the quark and near lepton can
have either the same or opposite helicity. We will follow the
conventions of \cite{smillie_old,smillie,Athanasiou:2006hv,Miller:2005zp} and label these
\begin{itemize}
\item Process 1: $\{q,l^{near},l^{far}\} = \{q_L,(l^-)_L,(l^+)_L
\}$ or $\{(\bar{q})_L,(l^+)_L,(l^-)_L \}$ or
$\{q_L,(l^+)_R,(l^-)_R \}$ or $\{(\bar{q})_L,(l^-)_R,(l^+)_R \}$;
\item Process 2: $\{q,l^{near},l^{far}\} = \{q_L,(l^+)_L,(l^-)_L
\}$ or $\{(\bar{q})_L,(l^-)_L,(l^+)_L \}$ or
$\{q_L,(l^-)_R,(l^+)_R \}$ or $\{(\bar{q})_L,(l^+)_R,(l^-)_R \}$.
\end{itemize}

A note regarding the quark's chirality is in order. In our SUSY
scenario, since $\tilde{\chi}^0_2$ is mostly wino, the decay
$\tilde{t}/\tilde{b} \rightarrow t/b \, \tilde{\chi}^0_2$ projects
out the LH part of the squark, which is a large mixture of both
chiral eigenstates (see section~\ref{mass_spectrum}). On the other
hand, for UED, it is possible that the mass eigenstates of the
third generation KK-quarks would be almost identical to the chiral
eigenstates, and $Z^*$ (defined in table~\ref{ued_spec}) tends to
be mostly $W^{3*}$ \cite{kk_masses}. Therefore, given that the
specific decay chain discussed here is observed, the LH KK-quark
(which is the heavier state~-- $b_2^*$ and $t_2^*$ in
table~\ref{ued_spec}) would be the main contributor. This is also
the case for first and second generation squarks in SUSY. As a
result, we assume in any case that the quark is left-handed. Note
that in order for an asymmetry to be produced (see definition in
the next section), the particles should have a well-defined
chirality either due to coupling or due to spectrum.

The matrix elements (and angular distributions) can be expressed
in terms of the masses of the particles and the three invariant
masses of the quark plus near lepton, the quark plus far lepton,
and the dilepton. This has been performed for all the possible
processes and spin assignments in \cite{smillie,Burns:2008cp} (and
also partially in \cite{Miller:2005zp}), assuming the
approximation in which all SM particles are massless and
intermediate particles have a zero width. Here we quote the
results relevant for our discussion.

The two most useful invariant mass distributions are for the quark
plus near lepton ($m_{ql}^{near}$) and the quark plus far lepton
($m_{ql}^{far}$). The former is given by \cite{smillie}
\begin{equation}
(m_{ql}^{near})^2=\frac{1}{2}(1-x)(1-y)(1-\cos \theta ^*)m^2_D ,
\end{equation}
where $\theta ^*$ is the angle between the quark and the near
lepton, in the rest frame of particle C. The rescaled invariant mass
is defined to be
\begin{equation} \label{qlnear_def}
\hat{m}_{ql}^{near} \equiv m_{ql}^{near}/
(m_{ql}^{near})_{\mathrm{max}}=\sin ( \theta ^* /2).
\end{equation}
The invariant mass distribution depends on the model, and changes
between the two processes defined above. For the SUSY case, the
distributions for the two processes, normalized to unit area, are
given by
\begin{equation} \label{qlnear_susy}
\frac{dP_1}{d(\hat{m}_{ql}^{near})^2}=2m^2 \, , \qquad
\frac{dP_2}{d(\hat{m}_{ql}^{near})^2}=2(1-m^2).
\end{equation}
The distributions for the UED scenario are
\cite{smillie,Burns:2008cp}
\begin{equation}
\begin{split}
\frac{dP_1}{d(\hat{m}_{ql}^{near})^2}&= \frac{3}{(1+2x)(2+y)} \left[
y+4(1-y+xy)m^2-4(1-x)(1-y)m^4 \right] \, , \\
\frac{dP_2}{d(\hat{m}_{ql}^{near})^2}&= \frac{3}{(1+2x)(2+y)} \left[
4x+y+4(1-2x-y+xy)m^2-4(1-x)(1-y)m^4 \right].
\end{split}
\end{equation}

The invariant mass for the quark and the far lepton is given by a
more complicated expression
\begin{equation}
\begin{split}
(m_{ql}^{far})^2=\frac{1}{4}(1-x)(1-z) & \left[ (1+y)(1-\cos \theta
\cos \theta ^*)+(1-y)(\cos \theta ^* -\cos \theta) \right. \\&\left.
-2 \sqrt{y} \sin \theta \sin \theta ^* \cos \phi \right] m^2_D \, ,
\end{split}
\end{equation}
where $\theta ^*$ is as before, $\theta$ is the angle between the
two leptons in the rest frame of particle B and $\phi$ is the angle
between the $ql^{near}$ and the dilepton planes, in the rest frame
of B. The rescaled invariant mass is defined as
\begin{equation}
\begin{split}
\hat{m}_{ql}^{far} \equiv m_{ql}^{far}/
(m_{ql}^{far})_{\mathrm{max}}= \frac{1}{2} & \left[ (1+y)(1-\cos
\theta \cos \theta ^*)+(1-y)(\cos \theta ^* -\cos \theta) \right.
\\&\left. -2 \sqrt{y} \sin \theta \sin \theta ^* \cos \phi \right]
^{\frac{1}{2}}.
\end{split}
\end{equation}
The invariant mass distributions for the SUSY scenario are
\cite{smillie,Burns:2008cp,Miller:2005zp}
\begin{equation} \label{qlfar_susy}
\begin{split}
\frac{dP_1}{d(\hat{m}_{ql}^{far})^2}&= \frac{-2}{(1-y)^2} \left\{
\begin{array}{ll} (1-y+\log y) & 0 \leq m^2 \leq y \\ (1-m^2+\log
m^2) & y<m^2 \leq 1 \end{array} \right. \\
\frac{dP_2}{d(\hat{m}_{ql}^{far})^2}&= \frac{2}{(1-y)^2} \left\{
\begin{array}{ll} (1-y+y \log y) & 0 \leq m^2 \leq y \\ (1-m^2+y \log
m^2) & y<m^2 \leq 1 \, , \end{array} \right.
\end{split}
\end{equation}
and for the UED case \cite{smillie,Burns:2008cp}
\begin{equation}
\begin{split}
\frac{dP_1}{d(\hat{m}_{ql}^{far})^2}&=
\frac{6}{(1+2x)(2+y)(1+2z)(1-y)^2} \times \\
& \left\{ \begin{split} (1-y) \left[ 4x-y+2z(2+3y-2x(5+y))-4m^2
(2-3x)(1-2z) \right]& \\ -\left[ y(1-2z(4+y))+4x(2z-y(1-4z))+4m^2
(1+y-x(2+y\right.&\left.))(1-2z) \right] \log y \\ & 0 \leq m^2 \leq y \\
(1-m^2) \left[ 4x(1+2y-5z-6yz)-5y+2z(2+9y)-4m^2 (1-x)\right.&\left.(1-z) \right] \\
-\left[ y(1-2z(4+y))+4x(2z-y(1-4z))+4m^2
(1+y-x(2+y\right.&\left.))(1-2z) \right] \log m^2 \\ & y<m^2 \leq 1
\end{split} \right. \\
\frac{dP_2}{d(\hat{m}_{ql}^{far})^2}&=
\frac{6}{(1+2x)(2+y)(1+2z)(1-y)^2} \times \\
& \left\{ \begin{split} (1-y) \left[ -y+2z(2+3y+2x(1-y))-4m^2
(2-x)(1-2z) \right]& \\ -\left[ y(1-2z(4+y))+4m^2
(1+(1-x)y)(1-2z) \right] \log y & \\ & 0 \leq m^2 \leq y \\
(1-m^2) \left[ 4z(1+x)-y(5-18z+8xy)-4m^2 (1-x)(1-z) \right] & \\
-\left[ y(1-2z(4+y))+4m^2 (1+(1-x)y)(1-2z) \right] \log m^2 & \\ &
y<m^2 \leq 1 \, . \end{split} \right.
\end{split}
\end{equation}

It should be noted that all the above distributions are given for
vanishing masses of the SM particles. This assumption is obviously
incorrect for the case of the top quark. The main effect of the
top mass would be a shift in the edges of the distributions, that
is, a change in the minimal and maximal possible value of
$m_{ql}^2$ (see appendix \ref{top_correction}). Other than that,
the form of the distributions is similar, as shown in
section~\ref{mc}.

\section{Experimentally-Accessible Distributions}
\label{exp_distributions}

As noted in \cite{barr}, there are experimental difficulties in
making a direct measurement of the above distributions, since the
two leptons cannot be easily identified as the near or far
leptons, and the charge of the quark is difficult to determine.
Hence, the distributions for processes 1 and 2 get mixed up. The
approach proposed in \cite{barr} uses the nature of the LHC, which
tends to produce more quarks than anti-quarks, as a mean to
establish an asymmetry between the two processes. A calculation
using an event generator yields an asymmetry of about 70\% for all
the relevant mass spectra, based on the proton PDF \cite{smillie}.

The most comprehensive study of this form was recently performed
in \cite{Burns:2008cp}, where the analysis was generalized
model-independently and efficiently represented in terms of the
useful observables. However, a physical source of asymmetry is
still required for a clear distinction between different spin
configurations.

The suggestion discussed here is to determine the charge of the
quark, in case it is a bottom or a top, in order to induce such an
asymmetry between the two oppositely-contributing processes. The
bottom decays through a $W$ boson to produce a lepton at a
probability of about 20\%. The sign of the lepton can be used to
determine the bottom's charge. Similarly, the top quark always
decays to a bottom and a $W$ boson, and when the $W$ decays
leptonically, the charge of the top can be identified. Obviously,
the production asymmetry of the LHC is irrelevant for third
generation quarks.

There is, however, one complication regarding the bottom: After it
is hadronized, it might take part in charge-flipping processes,
that is $B \to D$ meson decay or $B^0-\overline{B^0}$
oscillations. As a result, there would be a 30\% mistag rate
\cite{top_charge}. Fortunately, this is not the case for the top
quark, where the mistag rate is much lower, which is negligible
for our needs.

Despite the fact that the asymmetry which stems from the $pp$
collider is accidently of the same order as the bottom charge-id
efficiency, there is one crucial advantage in the latter approach.
The size of the production asymmetry is estimated in a numerical
calculation using an extrapolation of the PDF. On the other hand,
the bottom charge mistag rate is based on simple well-known
physical phenomena, which can be calculated from first principles,
and in fact was accurately measured.

The treatment of the distributions here is similar to
\cite{smillie} (as mentioned above, a more general representation
of the analysis was given in \cite{Burns:2008cp}, but for our
purpose this should suffice). Assuming that a quark was measured
(not an anti-quark) and that the leptons are right-handed (see
discussion at the end of this section), we consider the two
experimentally-accessible distributions for $ql^{\pm}$
\begin{equation} \label{obs_dist}
\begin{split}
\frac{dP}{d \hat{m}_{ql^-}^2}&=\frac{1}{2} \left[ f_c \left(
\frac{dP_2}{d(\hat{m}_{ql}^{near})^2}
+\frac{dP_1}{d(\hat{m}_{ql}^{far})^2} \right) +f_m \left(
\frac{dP_1}{d(\hat{m}_{ql}^{near})^2}
+\frac{dP_2}{d(\hat{m}_{ql}^{far})^2} \right) \right] \\
\frac{dP}{d \hat{m}_{ql^+}^2}&=\frac{1}{2} \left[ f_c \left(
\frac{dP_1}{d(\hat{m}_{ql}^{near})^2}
+\frac{dP_2}{d(\hat{m}_{ql}^{far})^2} \right) +f_m \left(
\frac{dP_2}{d(\hat{m}_{ql}^{near})^2}
+\frac{dP_1}{d(\hat{m}_{ql}^{far})^2} \right) \right],
\end{split}
\end{equation}
where $f_c$ is the probability for correctly identifying the
quark's charge and $f_m$ is the mistag rate ($f_c+f_m=1$). As
mentioned before, for a $b$-quark $f_c=0.7$, while for a top we
take $f_c=1$. If either an anti-quark is measured or the leptons
are left-handed, the expressions above are interchanged (and in
the case of both an anti-quark and LH leptons, the expressions
should not be interchanged). Notice that $m_{ql}^{near}$ and
$m_{ql}^{far}$ do not have the same range of possible values, so
the normalization of $m_{ql}^{\pm}$ is taken relative to the
maximal value of $m_{ql}$ in each case.

\begin{figure}[phbt]
\centering \subfloat[]{
\includegraphics[width=3.2In]{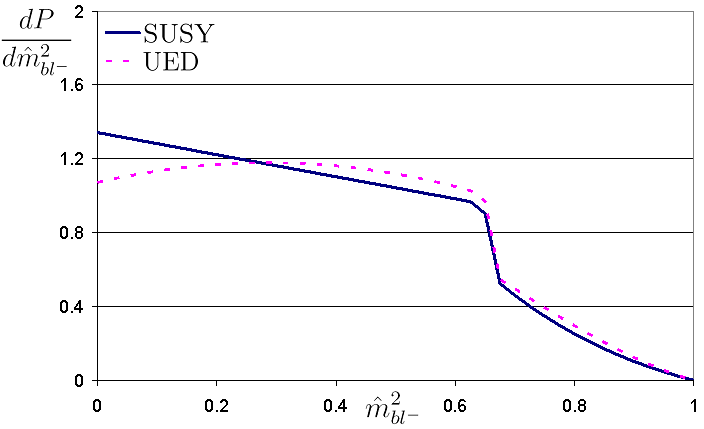}} \hspace{0.1In}
\subfloat[]{ \includegraphics[width=3.2In]{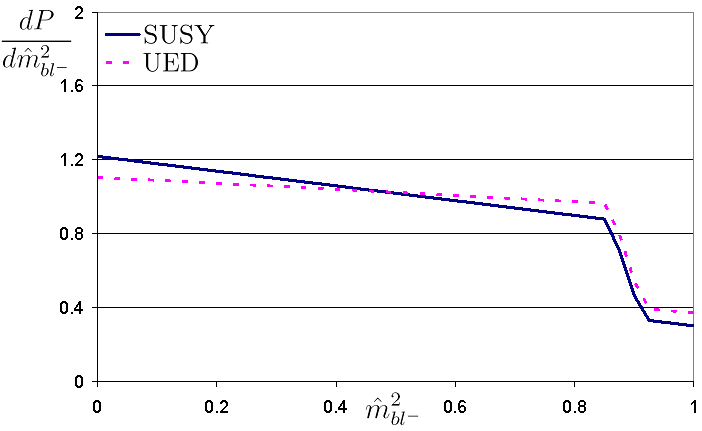}}
\caption{$b+l^-$ invariant mass distribution for (a) SUSY SPS1a
mass spectrum with a $\tilde{b}_1$ squark (see
table~\ref{susy_spec}) (b) UED mass spectrum with a $b^*_2$
KK-quark (table~\ref{ued_spec}). 30\% bottom mistag rate is
assumed. $\hat{m}_{bl^-}$ is normalized as $\hat{m}_{bl}^{far}$.}
\label{fig:blminus}
\end{figure}

\begin{figure}[phbt]
\centering \subfloat[]{
\includegraphics[width=3.2In]{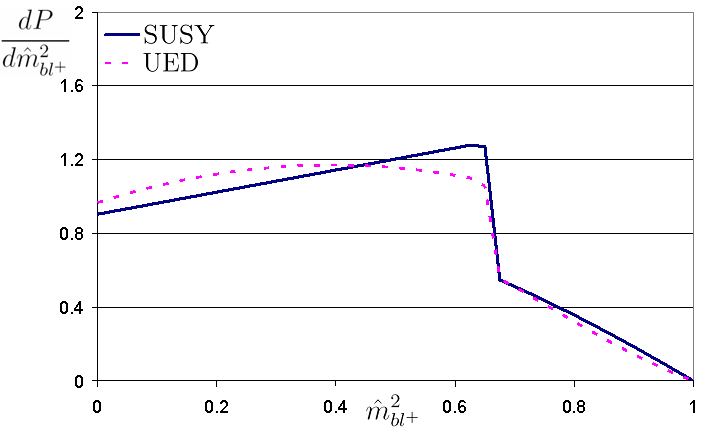}} \hspace{0.1In}
\subfloat[]{ \includegraphics[width=3.2In]{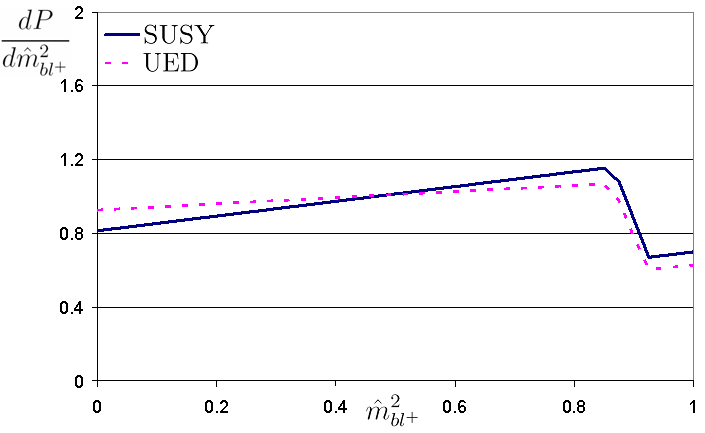}}
\caption{$b+l^+$ invariant mass distribution for (a) SUSY SPS1a
mass spectrum with a $\tilde{b}_1$ squark (b) UED mass spectrum
with a $b^*_2$ KK-quark. 30\% bottom mistag rate is assumed.}
\label{fig:blplus}
\end{figure}

\begin{figure}[phbt]
\centering \subfloat[]{
\includegraphics[width=3.2In]{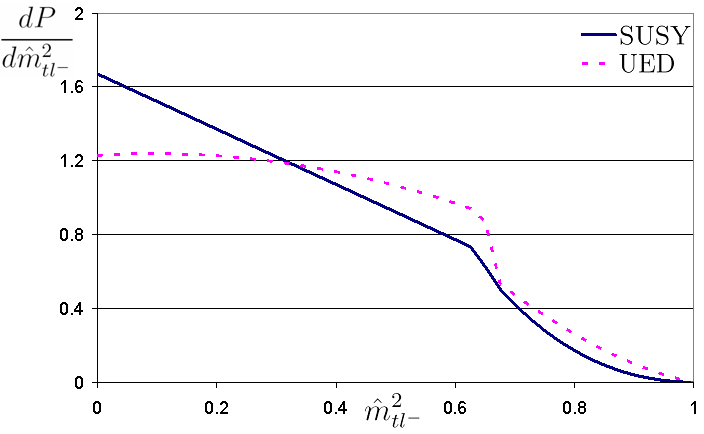}} \hspace{0.1In}
\subfloat[]{ \includegraphics[width=3.2In]{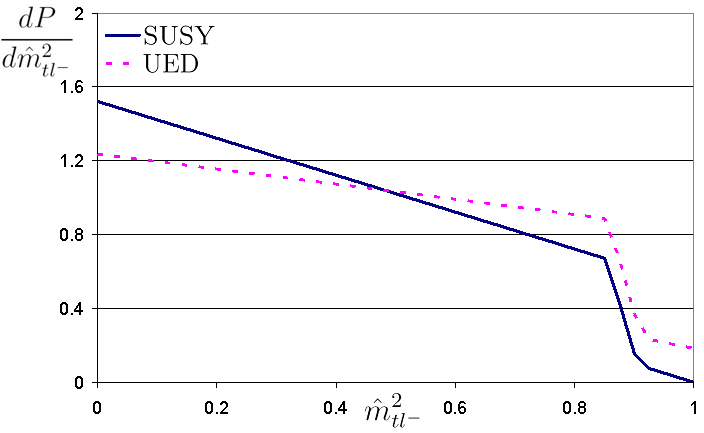}}
\caption{$t+l^-$ invariant mass distribution for (a) SUSY SPS1a
mass spectrum with a $\tilde{t}_1$ squark (b) UED mass spectrum
with a $t^*_2$ KK-quark.} \label{fig:tlminus}
\end{figure}

\begin{figure}[phbt]
\centering \subfloat[]{
\includegraphics[width=3.2In]{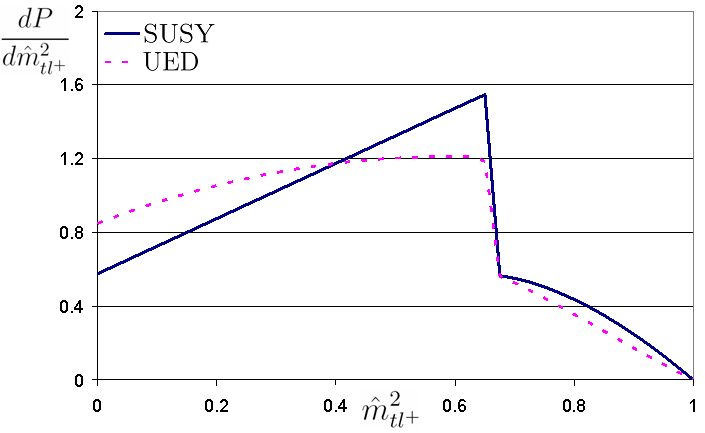}} \hspace{0.1In}
\subfloat[]{ \includegraphics[width=3.2In]{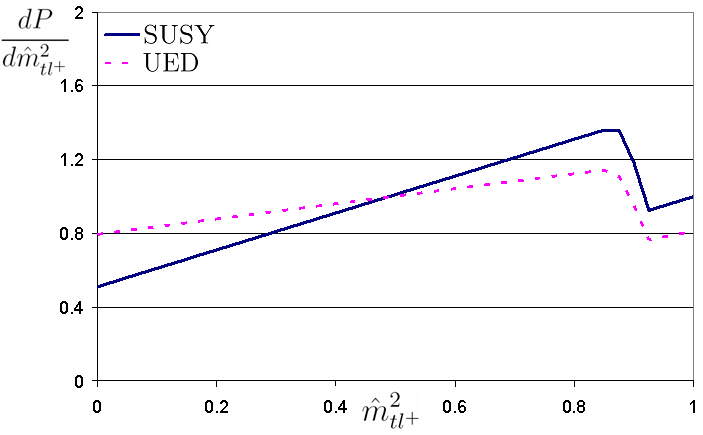}}
\caption{$t+l^+$ invariant mass distribution for (a) SUSY SPS1a
mass spectrum with a $\tilde{t}_1$ squark (b) UED mass spectrum
with a $t^*_2$ KK-quark.} \label{fig:tlplus}
\end{figure}

Figs.~\ref{fig:blminus} and \ref{fig:blplus} show a comparison of
the distributions for the SUSY and UED scenarios for the case of
the bottom quark (with 30\% mistag rate). Figs.~\ref{fig:tlminus}
and \ref{fig:tlplus} depict a comparison for the case of the top
quark. We focus for now on the light squark states for SUSY,
$\tilde{b}_1$ and $\tilde{t}_1$, and the heavy states for UED,
$b^*_2$ and $t^*_2$, since these are mostly left-handed (see
section~\ref{mass_spectrum} and remark in
section~\ref{Angular_Distributions}).

\begin{figure}[phbt]
\centering \subfloat[]{
\includegraphics[width=3.2In]{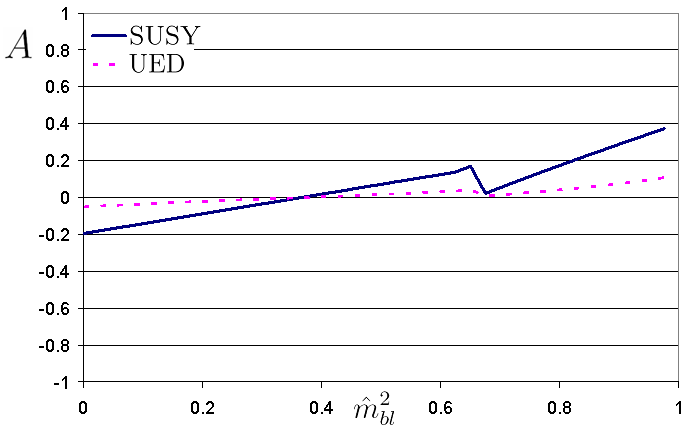}} \hspace{0.1In}
\subfloat[]{ \includegraphics[width=3.2In]{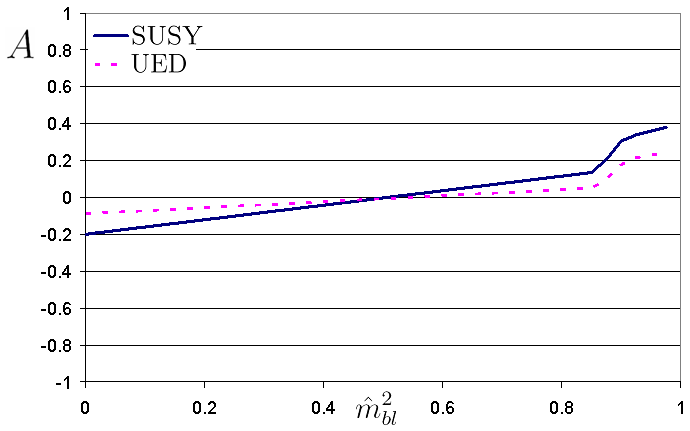}}
\caption{Asymmetry $A$ with a $b$-quark for (a) SUSY SPS1a mass
spectrum with a $\tilde{b}_2$ squark (b) UED mass spectrum with a
$b^*_1$ KK-quark. 30\% bottom mistag rate is assumed.}
\label{fig:asymb}
\end{figure}

\begin{figure}[phbt]
\centering \subfloat[]{
\includegraphics[width=3.2In]{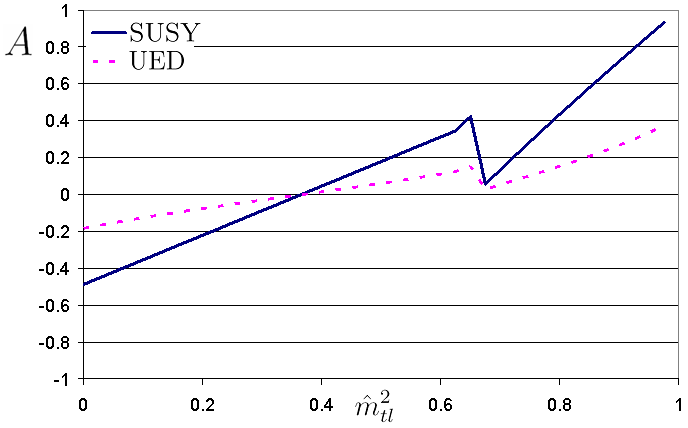}} \hspace{0.1In}
\subfloat[]{ \includegraphics[width=3.2In]{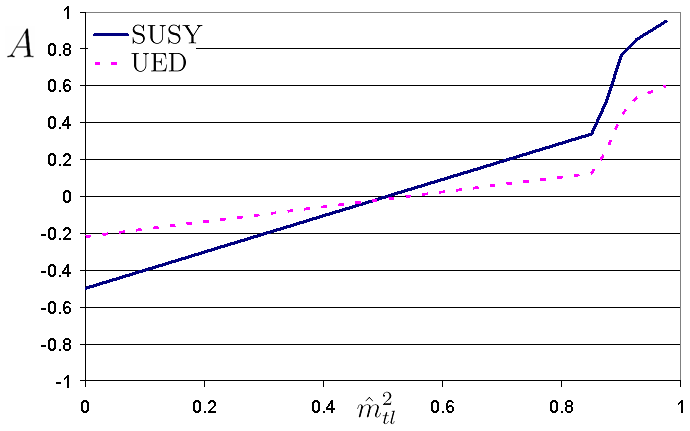}}
\caption{Asymmetry $A$ with a $t$-quark for (a) SUSY SPS1a mass
spectrum with a $\tilde{t}_1$ squark (b) UED mass spectrum with a
$t^*_2$ KK-quark.} \label{fig:asymt}
\end{figure}

Similar to \cite{barr,smillie_old,smillie}, it is useful to define
the charge asymmetry
\begin{equation}
A \equiv \frac{dP/d \hat{m}_{ql^+}^2-dP/d \hat{m}_{ql^-}^2}{dP/d
\hat{m}_{ql^+}^2+dP/d \hat{m}_{ql^-}^2}.
\end{equation}
A comparison of the asymmetry between SUSY and UED is given in
Figs.~\ref{fig:asymb} and \ref{fig:asymt}.

According to these plots, it seems easier to determine the spin
configuration in the top case than for the bottom case. This is a
direct consequence of the assumption that top's charge can be
measured more accurately, with no significant mistagging.

Comparing these results to the ones in Ref.~\cite{smillie}, we see
that the asymmetry in the bottom case is almost identical, since
the only difference lies within the mass of particle
D\footnote{Actually, for the SUSY spin configuration the results
are completely identical, since the form of the distributions is
independent of the squark's mass, see Eq.~\eqref{qlnear_susy} and
\eqref{qlfar_susy}, while for UED spins, the difference is very
small.}. Therefore, the improvement in our suggestion is
threefold:
\begin{itemize}
\item It allows accounting for third generation quarks to enhance
the statistics of the method. \item The evaluation of the 30\%
mistag rate for the bottom is more robust than the estimation of
the quark/anti-quark production asymmetry based on the PDF. \item
It produces a more distinguishable asymmetry for top quarks.
\end{itemize}
On the down side, the final states within this method are more
complicated, as they involve the quark's decay products, and the
cross section (even though model-dependent) tends to be smaller.

A few comments are in order here. First, a possible difficulty in
making the required measurement is to reconstruct the quark's
4-momentum, in case it is a top. This issue is discussed in
appendix \ref{top_reconstruct}. We propose a way to circumvent
this possible problem in section~\ref{daughters}. An interesting
approach for missing momenta reconstruction was recently suggested
in~\cite{Cho:2008tj}. In our case, due to an additional source of
missing energy originating from the leptonic decay of the quark,
that method cannot be applied directly. However, it would be
interesting to investigate whether an appropriate generalization
of that method could be formulated.

Another issue that needs to be carefully treated is a possible
confusion in distinguishing between the two leptons in the main
decay chain~\eqref{decay_chain} and the lepton that comes from the
top decay, which is used to determine its charge (for the bottom
quark, the $b$-tagging should assist in associating the lepton
correctly, hence we assume that this issue is solvable). This is
discussed in appendix~\ref{lepton_id}. This appendix also briefly
discusses the issue of correctly identifying the bottom from the
``right'' side of the cascade.

Regarding the leptons' chirality, note that particle C
($\tilde{\chi}_2^0$) can only decay to right-handed sleptons for
SPS1a (see table \ref{susy_spec}). More generally, when both left-
and right-handed sleptons are present in the spectrum, the LH
state would usually be the preferred decay product, because of the
chiral nature of $\tilde{\chi}_2^0$. In any case, an equal
contribution from both states requires some fine tuning of the
mass spectrum, so we could generally expect that only one
chirality should be involved.

Finally, an interesting insight can be inferred from the plots
above. The overall slope of the asymmetry curves in
Figs.~\ref{fig:asymb} and \ref{fig:asymt} is positive. This relies
on the assumption that the emitted leptons should be right-handed,
for the SUSY spectrum that we investigate. If for some other
spectrum, the leptons would be mostly left-handed, then the
observable distributions defined in Eq.~\eqref{obs_dist} should be
interchanged, and the slope of the asymmetry would be reversed.
This means that if we measure the charge of the quark, then the
slope of the asymmetry curve tells us about the chirality of the
leptons~\cite{eboli,Goto:2004cpa}. In fact, in conjunction
with~\cite{perelstein}, the chirality of the stop can also be
measured, and therefore render this conclusion model-independent.

\section{Monte Carlo Simulations} \label{mc}

In this section we present results for the above distributions, as
calculated by Monte Carlo simulations. All the calculations were
performed using MadGraph/MadEvent
\cite{Maltoni:2002qb,Stelzer:1994ta,madgraph}. For simplicity, we
focus only on the SUSY SPS1a mass spectrum, for both spin
configurations.

\begin{figure}[phbt]
\centering \subfloat[]{
\includegraphics[width=3.1In]{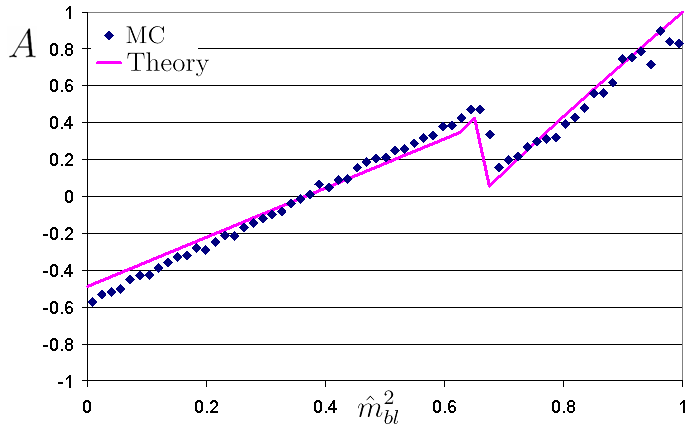}} \hspace{0.1In}
\subfloat[]{ \includegraphics[width=3.1In]{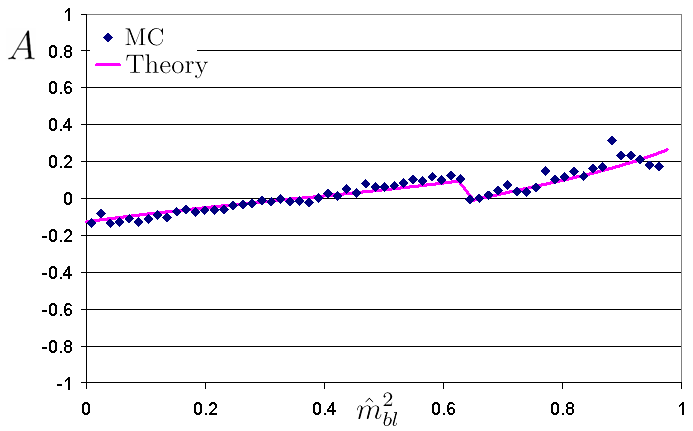}}
\caption{Theory vs.~MC comparison of the asymmetry $A$ with a
$b$-quark for (a)~SUSY (b)~UED. The mass of particle D is taken to
be that of $\tilde{b}_1$ of the SUSY spectrum. No mistagging is
assumed.} \label{fig:asymb_comp}
\end{figure}

First, we demonstrate the agreement between the theoretical
calculations described in previous sections and the MC results.
Fig.~\ref{fig:asymb_comp} shows a comparison of the asymmetry $A$
for a bottom quark.

\begin{figure}[phbt]
\centering \subfloat[]{
\includegraphics[width=3.1In]{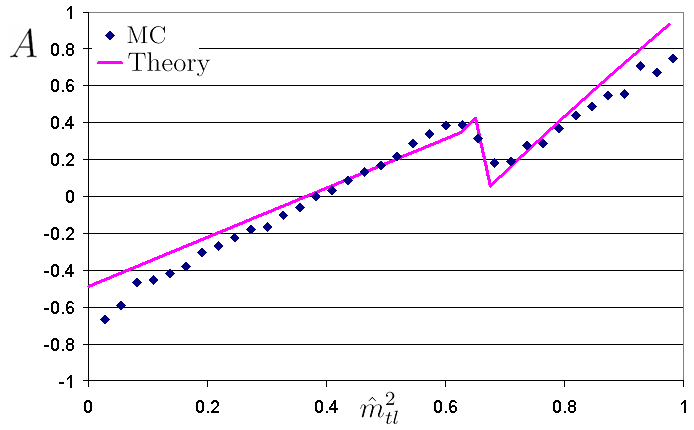}} \hspace{0.1In}
\subfloat[]{ \includegraphics[width=3.1In]{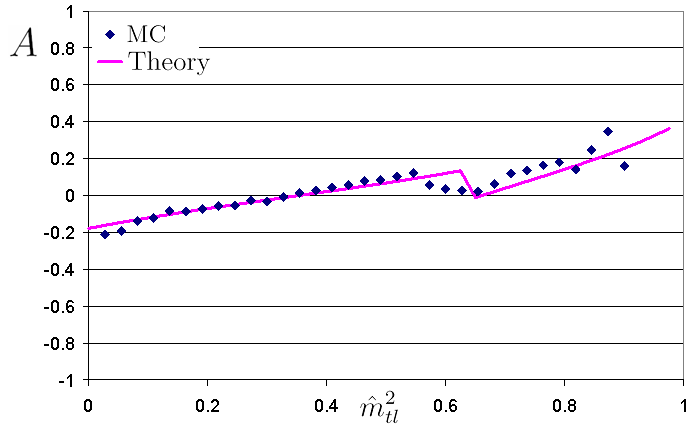}}
\caption{Theory vs.~MC comparison of the asymmetry $A$ with a
$t$-quark for (a)~SUSY (b)~UED. The mass of particle D is taken to
be that of $\tilde{t}_1$ of the SUSY spectrum.}
\label{fig:asymt_comp}
\end{figure}

In order to make this comparison for the top quark, some
adjustments are required, because the theoretical calculations in
the previous sections assumed a massless quark. As mentioned, the
mass of the top causes a shift in the minimal and maximal value of
$m_{ql}^2$ (see appendix~\ref{top_correction}). Hence, in this
comparison, we simply linearly normalize the values of $m_{ql}^2$
for the MC results to range from 0 to 1. As can be seen in
Fig.~\ref{fig:asymt_comp}, this is enough to yield a reasonable
agreement between theory and MC.

\begin{figure}[phbt]
\centering
\includegraphics[width=3.3In]{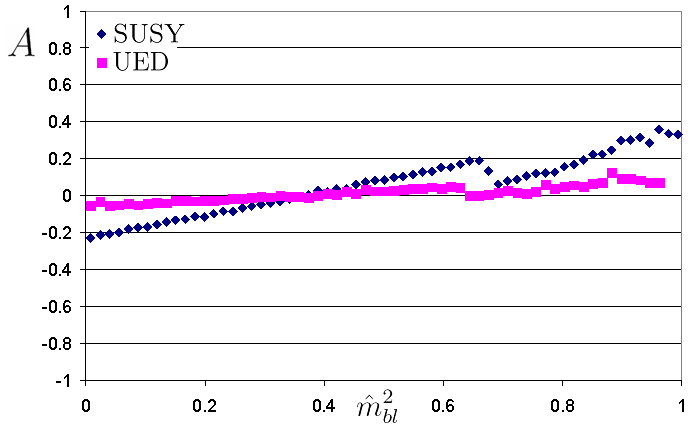}
\caption{Asymmetry $A$ with a $b$-quark. The mass of particle D is
taken to be that of $\tilde{b}_1$ of the SUSY spectrum. 30\%
bottom mistag rate is assumed.} \label{fig:b1}
\end{figure}

\begin{figure}[hptb]
\centering \subfloat[]{
\includegraphics[width=3.1In]{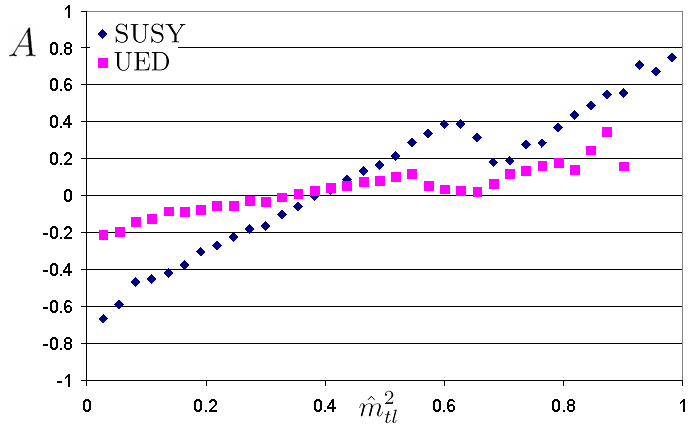}} \hspace{0.1In}
\subfloat[]{ \includegraphics[width=3.1In]{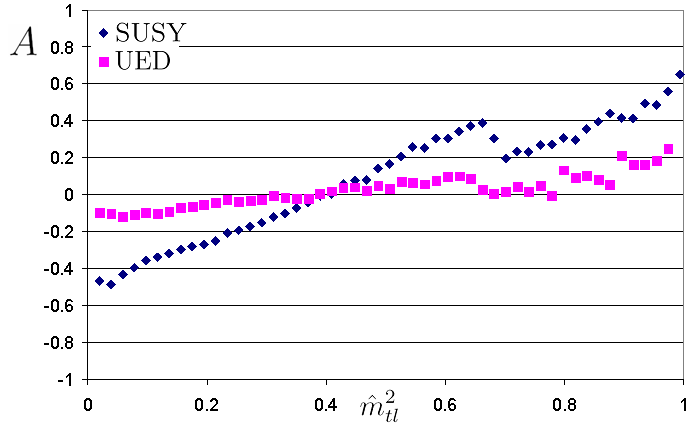}}
\caption{Asymmetry $A$ with a $t$-quark. The mass of particle D is
taken to be that of (a)~$\tilde{t}_1$ (b)~$\tilde{t}_2$ of the
SUSY spectrum.} \label{fig:t1t2}
\end{figure}

Next, we compare SUSY and UED distributions, in order to show that
our method allows us to distinguish between different models. The
asymmetry for a bottom and a top is shown in Figs.~\ref{fig:b1}
and \ref{fig:t1t2}, respectively. For the top we consider two
cases~-- one with a $\tilde{t}_1$ squark and one with
$\tilde{t}_2$. The first is the lighter state, so it is expected
to be more readily produced, while for the second state, the top
would be more highly-boosted, and therefore it is closer to the
massless limit. We again normalize $m_{ql}^2$ to range from 0 to
1.

We conclude this section by noting that a difference in the
asymmetry is manifested in the sbottom case, and even more so for
the stop.

\section{Distributions with the Top's Daughters} \label{daughters}

The top case presents a difficulty in making the experimental
measurement, since the reconstruction of the top's momentum might
be complicated. However, we expect that even the top's daughter
products will induce an asymmetry similar to the one obtained via
the top itself, as follows: In the limit of large boost, the decay
product tends to collimate and point roughly in the direction of
the top's boost, hence the angular distribution is hardly
affected. Furthermore, as was shown
in~\cite{perelstein,Almeida:2008tp}, the hardness of the $p_T$ of
both the lepton and the $b$-quark can serve as a top spin
analyzer.

\begin{figure}[hbt]
\centering \subfloat[]{
\includegraphics[width=3.1In]{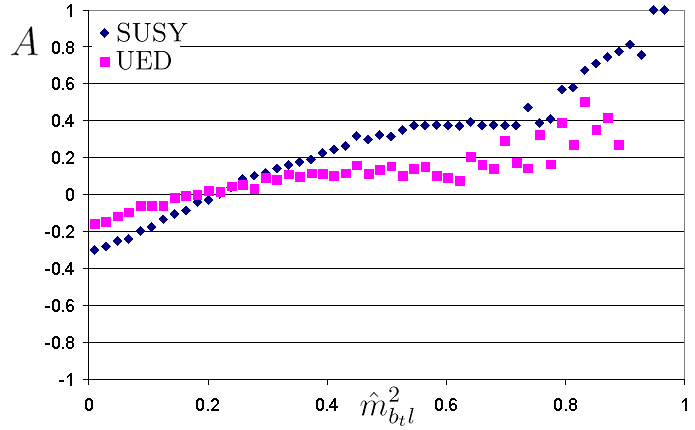}} \hspace{0.1In}
\subfloat[]{ \includegraphics[width=3.1In]{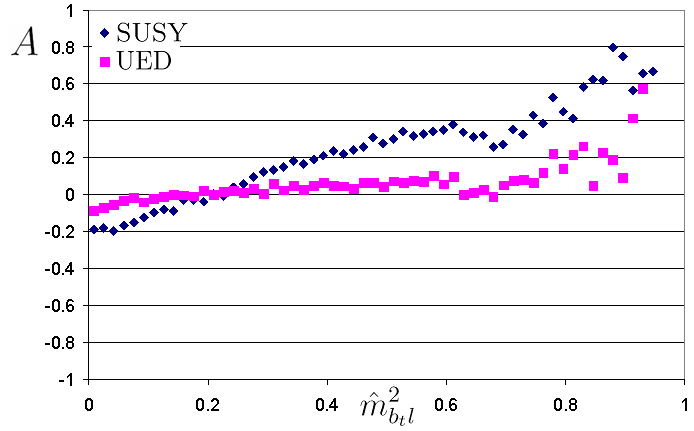}}
\caption{Asymmetry $A$ with the bottom which results from the top
decay ($b_t$). The mass of particle D is taken to be that of (a)
$\tilde{t}_1$ (b) $\tilde{t}_2$ of the SUSY spectrum.}
\label{fig:t1t2b}
\end{figure}

\begin{figure}[hbt]
\centering \subfloat[]{
\includegraphics[width=3.1In]{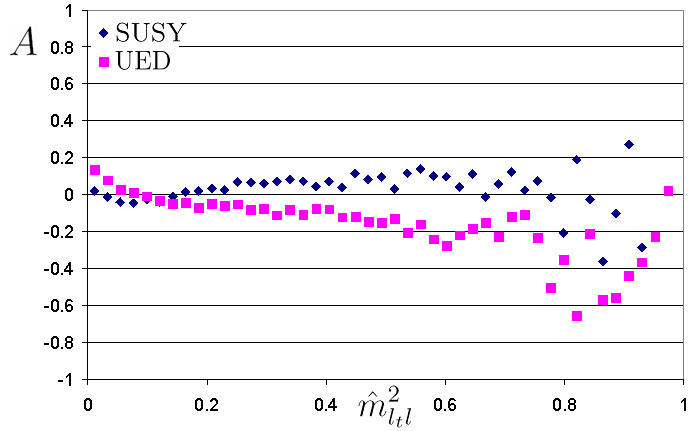}} \hspace{0.1In}
\subfloat[]{ \includegraphics[width=3.1In]{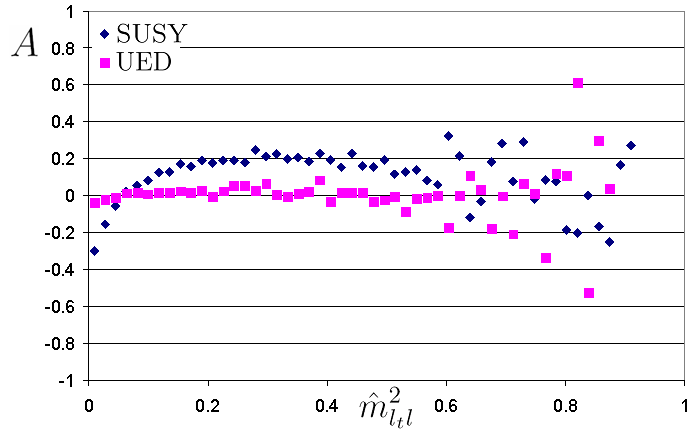}}
\caption{Asymmetry $A$ with the lepton which results from the top
decay ($l_t$). The mass of particle D is taken to be that of (a)
$\tilde{t}_1$ (b) $\tilde{t}_2$ of the SUSY spectrum.}
\label{fig:t1t2l}
\end{figure}

In fact, in~\cite{Almeida:2008tp} it was shown that for LH tops
(as in our case) the $b$-quark momentum tends to be harder, and
therefore it serves as a more reliable ``directionality''
analyzer, which is required to improve the quality of the
asymmetry. On the other hand, the lepton tends to decay in the
backward direction (opposite to the boost axis) for LH tops, and
therefore has a weaker correlation with the top's original
directionality.

\begin{figure}[hbt]
\centering \subfloat[]{
\includegraphics[width=3.1In]{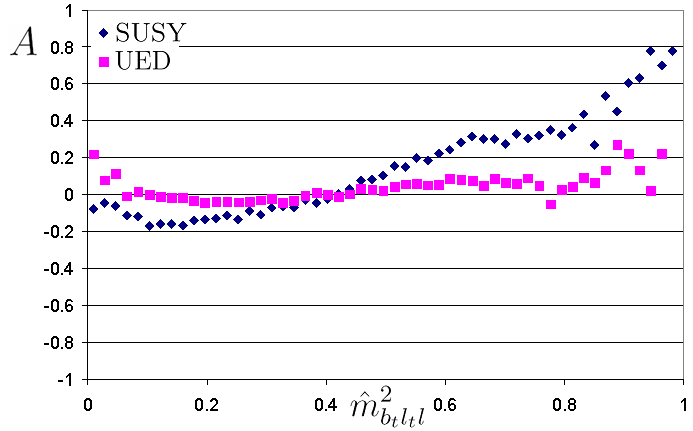}} \hspace{0.1In}
\subfloat[]{ \includegraphics[width=3.1In]{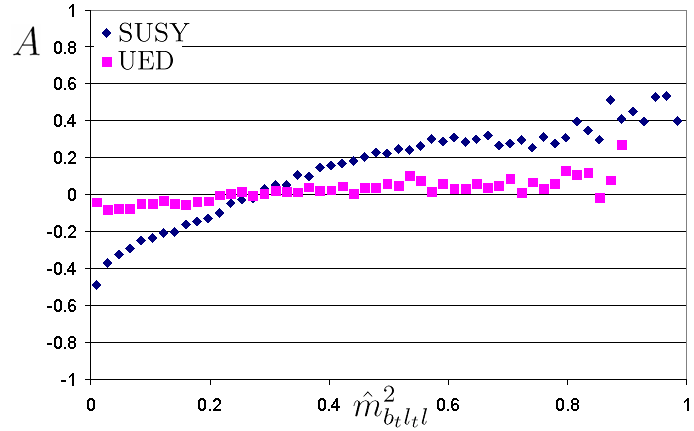}}
\caption{Asymmetry $A$ with the sum of the bottom and the lepton
which result from the top decay. The mass of particle D is taken
to be that of (a) $\tilde{t}_1$ (b) $\tilde{t}_2$ of the SUSY
spectrum.} \label{fig:t1t2bl}
\end{figure}

The feature described above is evident in the MC results, where
the asymmetry induced by the bottom daughter (shown in
Fig.~\ref{fig:t1t2b}) looks more similar to the original top one
(Fig.~\ref{fig:t1t2}), compared to the one induced by the lepton
daughter (Fig.~\ref{fig:t1t2l}). Obviously, combining the momenta
of both lepton and bottom daughter particles is expected to yield
an even more reliable angular distribution, as indeed observed by
the simulation in Fig.~\ref{fig:t1t2bl}.

To summarize, it is possible to determine the spin configuration
by using the top's decay products. The difference between SUSY and
UED is clearer with the bottom, while with the lepton only the
middle part of the distributions might yield a visible difference
(The $l_t+l^{\pm}$ distributions fall off rapidly with
$\hat{m}^2$, so for large values of $\hat{m}^2$ the asymmetry is
``noisier''). In any case, both the bottom and the lepton will be
available for analysis in each event (up to the $b$-tag
efficiency), so that the information extracted using them can be
combined to give a better distinction.

\section{Model Discrimination} \label{discrimination}
So far we have graphically shown the qualitative difference
between the two kinds of models for various methods. In order to
give a more quantitative sense of how well we can distinguish one
model from the other, we adapt the model discrimination method
used in the section 4 of~\cite{smillie}. We use the parton-level
distributions discussed above, leaving the analysis of realistic
data to future research.

Distinguishing between two spin configurations, such that the
``wrong'' configuration $S$ can be disfavored relative to the
``right'' one $T$ by a factor of $R$, requires $N$ events, which
is the number needed to be evaluated here. More explicitly, $R$
quantifies our requirement:
\begin{equation}
R=\frac{p(T|N \textrm{ events from } T)}{p(S|N \textrm{ events
from } T)} \, ,
\end{equation}
and we take it to be 1000, following \cite{smillie}.

The \textit{Kullback-Leibler} distance \cite{Kullback} between two
distributions is defined as
\begin{equation}
\mathrm{KL}(T,S) \equiv \int \log \left( \frac{p(m|T)}{p(m|S)}
\right) p(m|T) \textrm{d}m \, ,
\end{equation}
where $p(m|T)$ is an invariant mass distribution (properly
normalized as a probability distribution) coming from spin
configuration $T$, and the distance is not symmetric since $T$ is
taken to be the ``correct'' model. If we assume equal
\textit{a~priori} probabilities for the realization of each model
in nature, then in the limit of large $N$:
\begin{equation}
N \sim \frac{\log R}{\mathrm{KL}(T,S)} \, .
\end{equation}

This calculation can be applied to the $q+l^+$ and~$q+l^-$
distributions\footnote{The dilepton distribution can also be taken
into account, but it was found to be not so useful for
discriminating between SUSY and UED models \cite{smillie}.} (but
not to the asymmetry $A$, which is not properly normalized), by
analytically calculating KL$(T,S)$. According to the results of
\cite{smillie}, about 1060 events are needed to reject same-spin
models in case SUSY is the underlying model or 1090 events in the
opposite case (both cases with SPS1a mass spectrum, which will be
assumed throughout this section), using each kind of distribution
\emph{separately}. A combined Monte-Carlo calculation, taking into
account information from all the distributions together, yielded
about 450 events to reject UED (470 in the opposite case).

For the method discussed here, the calculation is similar. In the
case of the bottom quark, the distributions are almost identical
to those considered in \cite{smillie}, where the only minor
difference lies within the mass of the squark. Therefore, about
1140 events are required for each distribution separately, if SUSY
is realized in nature and the squark in the decay chain is
$\tilde{b}_1$ (for the heavier $\tilde{b}_2$ the result is
essentially the same as for first generation squarks).

The case of the top quark seems much better, as the charge mistag
rate is assumed to be negligible. On the other hand, the top
itself might not be reconstructed, so we should use its daughters
for a robust analysis. For example, using our MC results for the
$b_t+l_t+l^+$ distribution for a $\tilde{t}_1$ squark, we obtain
that 220 events are needed to disfavor UED over SUSY.

All this is based on considering independently one kind of
distribution each time. Combining all the information
simultaneously is much more efficient, but the calculation of $N$
is difficult to perform. Here we use a naive analysis: Assume that
$\mathrm{KL}_i$ is the Kullback-Leibler distance for distribution
of type $i$ ($i=q+l^+, \, q+l^-$ etc.) and $N_i=\log R
/\mathrm{KL}_i$ is the number of events needed to distinguish
between different models \emph{independently}. We wish to achieve
a discriminating factor of $R$ using all the distributions
together, based on a lower number of events $N$. So we can choose
a reduced factor $R_i$ for each distribution, such that $\log
R_i=N \cdot \mathrm{KL}_i$ (for each $i$ separately). But there is
still freedom to fix the overall scale of the $R_i$'s (and $N$) by
$\prod_i R_i=R$, to obtain the desired overall discrimination
factor. Summing over $i$, $N \sum_i \mathrm{KL}_i=\sum_i \log
R_i$, we have
\begin{equation} \label{combined_discrimination}
N=\frac{\sum_i \log R_i}{\sum_i \mathrm{KL}_i}=\frac{\log
R}{\sum_i \mathrm{KL}_i}=\left( \sum_i \frac{\mathrm{KL}_i}{\log
R} \right) ^{-1}= \left( \sum_i \frac{1}{N_i} \right) ^{-1}.
\end{equation}
This gives a result which is not as good as the complete
calculation of \cite{smillie}, that is, the number of events
calculated using Eq.~\eqref{combined_discrimination} is a bit
higher, but it is sufficient for our purpose.

Combining the two useful distributions available in the
$\tilde{b}_1$ case, $b+l^+$ and $b+l^-$, we see that about 570
events are required to disfavor UED by a factor of 1000, and 580
events to reject SUSY if a same-spin model is in action. Relying
on $b_t+l_t+l^{\pm}$ distributions, we need 210 and 240 events to
disfavor UED for $\tilde{t}_1$ and $\tilde{t}_2$, respectively
(240 and 250 in the opposite case). Overall, for the most
promising case of $\tilde{t}_1$, we expect that an integrated
luminosity of $\sim$100 fb$^{-1}$ should be enough to obtain a
clear distinction, for SPS1a mass spectrum (based on the cross
section estimation in section~\ref{mass_spectrum}).

It is important to emphasize that the numbers presented here are
only meant to give a sense of the distinguishablity offered by our
method, as in practice a more realistic analysis is required, and
the results are model-dependent.

\section{Conclusion} \label{conclusion}

Recognizing the nature of a new physics model at the LHC is a
non-trivial task, since various frameworks tend to yield similar
signatures. By measuring the spin of newly-discovered particles,
we can assist in revealing a part of this nature.

In this paper we suggest an improvement to an existing method to
determine the spin of intermediate particles in decay chains. This
is applied by taking third generation quarks, for which we can
measure the charge, in order to establish observable non-trivial
mass distributions. We believe that it is a robust method, as
hadronic uncertainties related to the PDFs do not take part in the
asymmetries.

Moreover, since the top's charge can be more accurately determined,
it allows for a better spin distinction. However, there might be
some difficulty with the reconstruction of the top's momentum.
Therefore, we showed that its decay products can be used to obtain
information about the spin, and we believe that this should be a
more natural approach to apply, rather than numerically solving for
the kinematics. Another benefit is the fact that the third
generation squarks/same-spin partners are different in mass than the
first two generations, so that even if the spectrum turns out to be
almost degenerate, these states should be better distinguished. For
concreteness, we also evaluate the significance of our method for a
specific spectrum, and compare it to previous existing methods.

There are some drawbacks in the method suggested here. First, it
might turn out to be a challenging task to obtain enough
statistics to enable a clear distinction, especially due to the
dilution factor induced by the specific decay of the quark, which
is required for charge measurement. Another possible difficulty is
in correctly identifying the lepton coming from the top quark,
particularly if the stop/same-spin top partner is relatively
light. Finally, we emphasize that our study does not attempt to
provide a fully realistic analysis, but rather aim to propose a
new approach and its potential feasibility. Further investigation
is required in order to make it fully applicable, which includes,
among others, issues like uncertainties related to new particles
masses, combinatorial background and various detector effects.

\section*{Acknowledgements}
We thank Kyoungchul Kong, Hitoshi Murayama, Michele Papucci and
Itay Yavin for helpful discussions. We also thank Oscar Eboli for
useful discussions and for commenting on the manuscript. The
research of GP is supported by PHY-06353354 NSF grant and the
Peter and Patricia Gruber Award.


\appendix
\section{Invariant Mass Corrections Due to the Top Mass} \label{top_correction}

The expressions for the invariant masses given in section
\ref{Angular_Distributions} were computed assuming that the quark's
mass vanishes. However, in the case of the top quark, its mass can
no longer be neglected. As a result, the distributions would have a
different maximum value for $m_{ql}^2$, and also a minimum which is
larger than 0. Here we do not calculate the full form of the
distributions, but only give the correction for the edges.

The maximum and minimum values for $(m_{tl}^{near})^2$ are given by
\begin{equation}
(m_{tl}^{near})^2_{\mathrm{max/min}}=m_D^2 \left[ t +\frac{1}{2}
(1-y) \left( 1-x-t \pm \sqrt{1+x^2+t^2-2x-2t-2xt} \right) \right]
\, ,
\end{equation}
where the $+$ sign before the square root is for the maximum
value, and the $-$ sign is for the minimum, and we defined $t
\equiv m_t^2/m_D^2$. This shifts up the maximum edge of the
distribution. For example in the case of a $\tilde{t}_1$ squark in
the SPS1a mass spectrum, the maximum changes from 216 to 248 GeV.
The minimum would be usually slightly above the top's mass~-- 187
GeV in this example.

Similarly, for $(m_{ql}^{far})^2$ we get
\begin{equation}
(m_{tl}^{far})^2_{\mathrm{max/min}}=m_D^2 \left[ t +\frac{1}{2}
(1-z) \left( 1-x-t \pm \sqrt{1+x^2+t^2-2x-2t-2xt} \right) \right]
\, .
\end{equation}

If we use the bottom that results from the top decay, instead of
the top itself (see section~\ref{daughters}), then the
distributions change. The minimum of $m^2$ is very close to 0 (or
the bottom's mass, to be exact), and the maximum for $b+l^{near}$
is
\begin{equation}
(m_{bl}^{near})^2_{\mathrm{max}}= \frac{1}{2} m_D^2(1-w)(1-y)
\left( 1-x-t +\sqrt{1+x^2+t^2-2x-2t-2xt} \right) \, ,
\end{equation}
where $w \equiv m_W^2/m_t^2$. For $b+l^{far}$ we simply replace $y
\rightarrow z$ in this formula, like before.

It should be mentioned that all the above formulae are still
slightly approximated, since we assumed that all intermediated
particles are on mass shell.

\section{Four-Momenta Reconstruction} \label{top_reconstruct}

The calculation of the invariant mass distributions requires the
knowledge of the 4-momenta of the relevant outgoing particles. In
the case of light quark production in the process, which is
observed as a jet, this should not be a problem. However, in the
scenario studied here, the quark might be a top which decays
leptonically ($t \to bW \to bl \nu$). Therefore, evaluating the
4-momentum of the top might be more difficult.

In order to demonstrate how this is possible, we now show some
examples. First, let's assume that a stop pair is produced, both
of which decay through the same process, Eq.~\eqref{decay_chain}.
Furthermore, we assume that one top quark decays leptonically, and
the other decays hadronically, so that the 4-momentum of the
latter can be reconstructed.

The number of unknown parameters is 12, which arises from the
4-momenta of one neutrino and two LSPs. The count of constraints
goes as follows: the mass of $\tilde{\chi}_1^0$ yields two
constraints (one on each side of the decay chain), the slepton
mass (2), the mass of $\tilde{\chi}_2^0$ (2), the neutrino mass
(1), the $W$-boson mass (1), the top mass (1), the stop mass (2)
and two missing $p_T$ constraints. Altogether, we have 13
equations satisfied by the 12 unknowns, which makes this process
over-constrained. We have verified numerically that a solution can
be obtained in this way (the excess constraint assists in
identifying the correct solution).

Another example is the case in which the second stop decays as
$\tilde{t} \to b \tilde{\chi}_1^+ \to b W^+ \tilde{\chi}_1^0$,
with a hadronic decay of the $W$-boson. Again, we have 12
unknowns, and the following constraints: the mass of
$\tilde{\chi}_1^0$ (2), the slepton mass (1), the mass of
$\tilde{\chi}_2^0$ (1), the neutrino mass (1), the $W$-boson mass
(1), the top mass (1), the mass of $\tilde{\chi}_1^+$ (1), the
stop mass (2) and two missing $p_T$ constraints. Hence, in this
case we have 12 constraints. This is, of course, valid for
truth-level analysis without taking into account detector effects
(for example, the effects of detector resolution and jet energy
scale) and other uncertainties such as new particles' masses in
the decay chain, etc.

In case of a shorter decay chain of the second stop, we might not
have enough information to reconstruct the missing 4-momenta,
hence a different approach is required (see
section~\ref{daughters}). We should also note that the production
of one stop and one gluino usually offers at least 12 constraints.

\section{Identifying Important Particles} \label{lepton_id}

There are several potential problems in correctly assigning labels
to particles in the decay chain. First, the issue of identifying
the lepton that should be used to determine the charge of the top
is certainly not a simple one. Naively, it might be expected that
if the top is highly-boosted, then this lepton should be more
collinear with the daughter bottom (however, note the discussion
in section~\ref{daughters} related to the difficulty in the case
of LH leptonic top). As a result, for the SUSY spectrum considered
here (table~\ref{susy_spec}), the heavier stop $\tilde{t}_2$ will
provide a greater chance to identify the lepton than the lighter
one $\tilde{t}_1$.

\begin{figure}[hbt]
\centering \subfloat[]{
\includegraphics[width=3.1In]{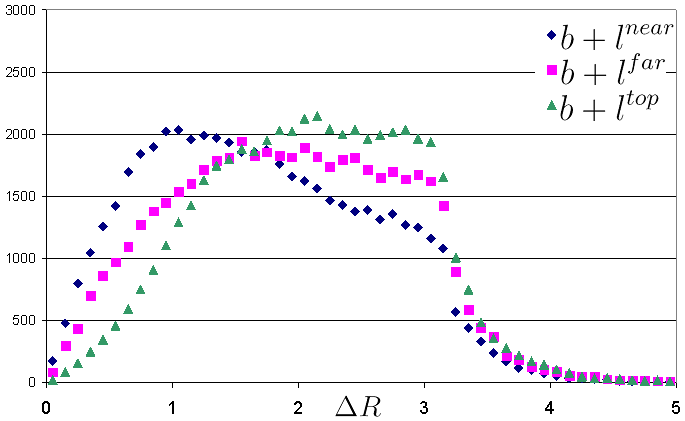}} \hspace{0.1In}
\subfloat[]{ \includegraphics[width=3.1In]{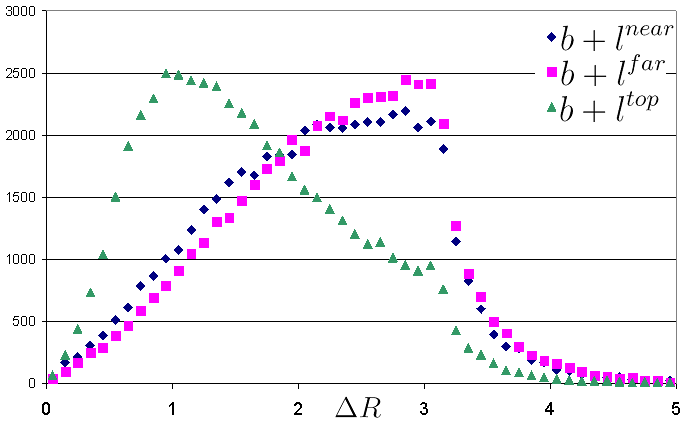}}
\caption{$\Delta R$ distribution of the bottom with each lepton
for (a)~$\tilde{t}_1$ (b)~$\tilde{t}_2$. The scale of the $y$-axis
is arbitrary. This assumes SUSY spin configuration and process 2,
but it is similar for both processes and spin setups. The mass
spectrum is SUSY SPS1a.} \label{fig:dr}
\end{figure}

This intuitive argument can be demonstrated by MC simulations.
Fig.~\ref{fig:dr} shows a comparison of $\Delta R$ of the
daughter bottom with each one of the three leptons, for both
stop mass eigenstates. It is evident that for $\tilde{t}_2$,
the lepton coming from the top decay does indeed tend to be
closer to the bottom than the other leptons, while for
$\tilde{t}_1$ that is not true.

\begin{figure}[hbt]
\centering \subfloat[]{
\includegraphics[width=3.1In]{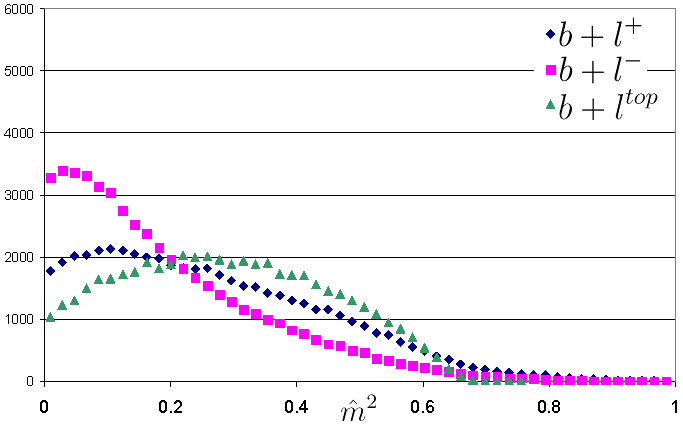}} \hspace{0.1In}
\subfloat[]{ \includegraphics[width=3.1In]{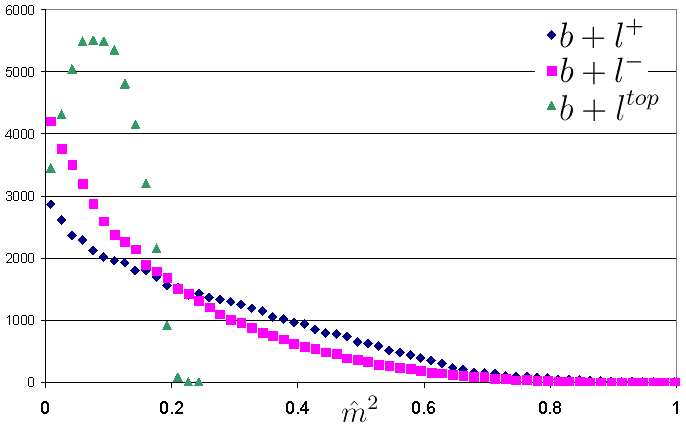}}
\caption{Invariant mass distribution of the bottom with each
lepton for (a)~$\tilde{t}_1$ (b)~$\tilde{t}_2$. The scale of the
$y$-axis is arbitrary. This assumes SUSY spin configuration and
sums over the two processes to obtain $b+l^{\pm}$ instead of
$b+l^{near/far}$ (the $b+l^{top}$ distribution is identical for
both processes). The mass spectrum is SUSY SPS1a.} \label{fig:mbl}
\end{figure}

Another variable we can use to differentiate between the
leptons is the invariant mass distributions of each lepton with
the daughter bottom. Since the invariant mass of the bottom and
the lepton from the top is bounded by the top mass, a cut on
this parameter can assist in this issue. As shown in
Fig.~\ref{fig:mbl}, this is again more useful for $\tilde{t}_2$
than for $\tilde{t}_1$.

\begin{figure}[hbt]
\centering \subfloat[]{
\includegraphics[width=3.1In]{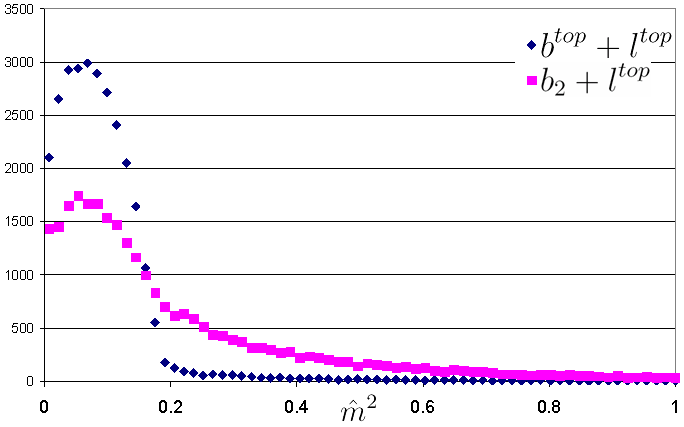}} \hspace{0.1In}
\subfloat[]{ \includegraphics[width=3.1In]{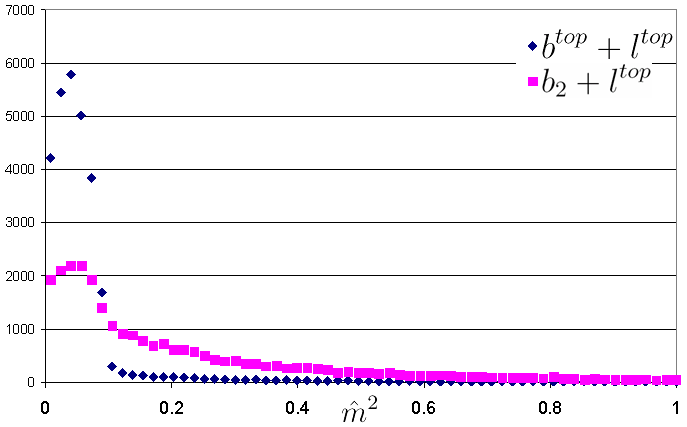}}
\caption{Invariant mass distribution of the bottom ($b^{top}$)
with the lepton from the same top (similar to the previous figure)
compared to the bottom from the other side of the cascade ($b_2$)
with the same lepton, for (a)~$\tilde{t}_1$ (b)~$\tilde{t}_2$. The
scale of the $y$-axis is arbitrary. This assumes SUSY spin
configuration and SPS1a mass spectrum.} \label{fig:mblbl}
\end{figure}

A similar issue arises with the bottoms. In any case of a third
generation cascade, there will be two bottom quarks from the
two sides of the cascade, and in general it might be difficult
to correctly assign the one that belongs to the ``interesting''
side. This issue can be approached in a similar fashion, for
instance by investigating each bottom's mass distribution with
a lepton from that side (assuming that the other side is
strictly hadronic). Fig.~\ref{fig:mblbl} shows that for a stop
event, the distribution of a bottom with a lepton coming from
the same top is different than the other bottom with the same
lepton, as the former is bounded by the top's mass.

Finally, we can use the approach discussed in the previous
appendix, and reconstruct the entire process. When this system
is over-constrained, the correct solution can be obtained.
Actually, this also makes it possible to identify correctly the
near and far leptons (under the above idealized assumptions).
In any case, this issue requires a more detailed study.

\bibliographystyle{utcaps}
\bibliography{spinbib}

\end{document}